\documentclass{cernrep} 
\usepackage{texnames}
\usepackage[T1]{fontenc}
\usepackage{mcite}

\begin{document}
\title{Real time description of fission}
 
\author{I. Stetcu$^1$, A. Bulgac$^2$, S. Jin$^2$, K. J. Roche$^{2,3}$, and N. Schunck$^{4}$}

\institute{$^1$Los Alamos National Laboratory, Los Alamos, New Mexico, 87545, USA \\
$^2$University of Washington, Seattle, Washington 98195-1560, USA \\
$^3$Pacific Northwest National Laboratory, Richland, WA 99352, USA, USA \\
$^4$Lawrence Livermore National Laboratory, Livermore,  California 94551, USA}

\maketitle 

\begin{abstract}
Using the time-dependent superfluid local density approximation, the dynamics of fission is investigated in real time from just beyond the saddle to fully separated fragments. Simulations produced in this fully microscopic framework can help to assess the validity of the current approaches to fission, and to obtain estimate of fission observables. In this contribution, we concentrate on general aspects of fission dynamics.
\end{abstract}
 
\section{Introduction}

 The microscopic description of nuclear fission remains a goal of nuclear theory even almost 80 years after its discovery. Recent developments, both in theoretical modeling and computational power, give us hope that progress can be finally made towards a microscopic theory of nuclear fission \cite{Schunck:2016}. Even if the complete microscopic description remains a computationally demanding task, the information that can be provided by current calculations can be extremely useful to guide and constraint more phenomenological approaches. First, a microscopic model that describes the real-time dynamics of the fissioning system can justify or rule out some of the approximations. Second, the microscopic approach can be used to obtain trends, e.g., with increasing excitation energy of the fissioning system, or even to compute observables that cannot be otherwise calculated in phenomenological approaches or that can be hindered by the limitations of the method. For example, in all phenomenological approaches, the full separation of the fragments cannot be modeled. While this approximation can have little impact on the mass numbers of the fission fragments, the same approximation can be of concern when the total kinetic energy (TKE) or the total excitation energy (TXE) of the fission fragments are computed, if these can be even computed in these approaches. Moreover, most phenomenological models implicitly built in the approximation that no neutrons are emitted at scission or during the acceleration of the fission fragments. Given that one cannot distinguish experimentally between neutrons emitted from the fission fragments after full acceleration, and neutrons emitted earlier in the process, only a microscopic theory able to follow in real time the evolution of the system to fully separated fission fragments can answer such questions.
 
Our approach to nuclear fission is  the time-dependent superfluid local density approximation (TDSLDA). This is an extension of the superfluid local density approximation (SLDA) introduced in 2002 as an alternative to the density functional theory (DFT) framework for superfluid systems of Oliveira, Gross, and Kohn~\cite{Oliveira:1988qy}, approach that lacked the local character and would be prohibitive to implement even on exascale computers. TDSLDA has becoome a very successful theoretical model. It reproduced correctly, and often predicted before experimental data became available,
a large number of phenomena and properties such as the ground state energy, pairing gap, collective modes, quantized vortices, see QMC and experimental  studies discussed in Refs.~\cite{Bulgac:2013b,Bulgac:2013d,Wlazlowski:2015,Bulgac:2011c}. In nuclear physics, TDSLDA was used to study the excitation of collective modes in deformed open shell nuclei, in particular triaxial nuclei, without any restrictions~\cite{Stetcu:2011}.  In the case of relativistic Coulomb excitation of Uranium~\cite{Stetcu:2014} the external field created by another impinging Uranium nucleus is so strong that nonlinear and non-adiabatic effects are very important, and cannot be captured by a traditional QRPA approach. The TDSLA incorporates the effects of the continuum, the dynamics of the pairing field, and the numerical solution is implemented with controlled approximations and with negligible numerical corrections~\cite{Bulgac:2013,Jin:2016}.

Within the TDSLDA we have investigated in a series of papers \cite{Bulgac:2015a,Bulgac:2018ukd} the fission of a $^{240}$Pu nucleus, following the dynamics of the process in real time from the outer saddle to scission and beyond. In our investigations, the fragments are well separated, which allows us to estimate quantities like the TKE or TXE and investigate the emission of neutrons at scission and from the fragments before full acceleration. We have shown that many collective degrees of freedom are excited in fission dynamics, on the way from saddle-to-scission, not only 2 or even 5, as used in other models~\cite{moller:2001,Sierk:2017}, and that the one-body dissipation plays an important role in the dynamics \cite{Bulgac:2018ukd}. We will discuss in this contribution some of the characteristics of the fission dynamics and discuss further developments of the method. The ultimate goal of the theoretical effort is to produce reliable description and/or trends of fission observables that can be used as input in applications.

\section{Theoretical Framework}

(TD)SLDA is formally equivalent to (TD)HFB or (TD) Bogoliubov-de Gennes mean-field approaches by design and it is a very complex mathematical problem. In various studies over the years  we solved up to 500,000 or more complex-valued, nonlinear, coupled TD partial differential equations (PDEs) on large 3D spatial lattices for up to $400,000$ of time steps, with very high numerical accuracy. This feat has only been feasible by using some of the largest supercomputers in the world with highly-optimized computer codes \cite{Bulgac:2008}. 

In TDSLDA, one assumes that the system is described by a single generalized Slater determinant, or HFB vacuum, composed of quasi-particle wavefunctions from the start to the finish of the simulations.  
The evolution of the wavefunctions is described by the time-dependent Schr\"odinger equation,
\begin{align} \label{eq:tdslda}
i\hbar \frac{\partial}{\partial t}
\begin{pmatrix}
u_{k\uparrow}  \\
u_{k\downarrow} \\
v_{k\uparrow} \\
v_{k\downarrow}
\end{pmatrix}
=
\begin{pmatrix}
h_{\uparrow \uparrow}  & h_{\uparrow \downarrow} & 0 & \Delta \\
h_{\downarrow \uparrow} & h_{\downarrow \downarrow} & -\Delta & 0 \\
0 & -\Delta^* &  -h^*_{\uparrow \uparrow}  & -h^*_{\uparrow \downarrow} \\
\Delta^* & 0 & -h^*_{\downarrow \uparrow} & -h^*_{\downarrow \downarrow} 
\end{pmatrix}
\begin{pmatrix}
u_{k\uparrow} \\
u_{k\downarrow} \\
v_{k\uparrow} \\
v_{k\downarrow}
\end{pmatrix},
\end{align}
where we have suppressed the spatial $\bm{r}$ and time coordinate $t$ and $k$  
labels the quasiparticle wavefunctions (qpwfs) $[u_{k\sigma}(\bm{r},t), v_{k\sigma}(\bm{r},t)]$, 
where $\sigma = \uparrow, \downarrow$.  The single particle Hamiltonian 
$h_{\sigma, \sigma'}(\bm{r}, t)$, and the pairing field $\Delta(\bm{r}, t)$ are 
functionals of various neutron and proton densities, which are determined 
by the quasiparticle wavefunctions, see Ref.~\cite{Jin:2016} for details. As input to the calculations we use Skyrme-type functionals, which ensure that the equations remain local and consistent with the Kohn-Sham philosophy. 
The TDSLDA equations (\ref{eq:tdslda}) are discretized and solved on rectangular lattices. The size  of the discretized Hamiltonian in Eq. \eqref{eq:tdslda} is $4 N_xN_yN_z\times 4 N_xN_yN_z$, 
where $N_x,\; N_y,\; N_z$ are the number of lattice points in the corresponding spatial directions. 
Each qpwf has 4 components and thus one has to solve $16N_xN_yN_z$ partial 
differential equations (PDEs), where each function is defined on  $N_xN_yN_z$ lattice points. 
Over the years, we have developed a highly efficient code which takes advantage of the GPU accelerators 
and which provide an enormous speedup with respect to a CPU-only code. On Titan a GPU code, 
a single trajectory from saddle to scission ($10^3$ to $10^4$ fm/c) can be finished 
within 12 hours using $1,000$ GPUs with a time step $\Delta t = 0.03~\mathrm{fm/c} $.

The main ingredient necessary in fission simulations, which also makes the numerical calculations much more complex than in other approaches,
is the pairing field in Eq. (\ref{eq:tdslda}). It was well understood long time ago that without including pairing correlations a nucleus 
  will not fission at low energies in a microscopic dynamic approach ~\cite{Bertsch:1987,Barranco:1988,Bertsch:1997}. In calculations performed in the time-dependent Hartree-Fock (TDHF) model, fission was obtained only by introducing unrealistically large pairing gaps \cite{Negele:1978}. In other more recent simulations like~\cite{Tanimura:2015,Hashimoto:2016,Goddard:2015,Goddard:2016}, the system fissioned only at high energy, or if the initial state was far along the fission path so that the two fragments were already formed.

Only one-body (current) densities are included in TDSLDA. Hence, while the system is described by a single generalized Slater determinant throughout the simulations, we observe the separation between the fission fragments by looking at the density profile. 
As noted in Ref. \cite{Bulgac:2015a}, the dynamics involves a large number of degrees of freedom, and can take a long time, although the duration of the process depends very strongly on the nuclear energy density functional (NEDF) used in the calculation. The results obtained in TDSLDA are consistent \cite{Bulgac:2015a} with expectations that the light fragment emerges deformed, while the heavy fragment is close to spherical shape, as it is expected to be close to a closed shell configuration.
Fission quantities such as the fission fragment masses, charges, kinetic and excitation energies \etc, are calculated after separation, by dividing the simulation box in two at the separation line.

\section{Selected features of the fission dynamics}

\begin{figure}[t]
\centering\includegraphics[width=.36\linewidth]{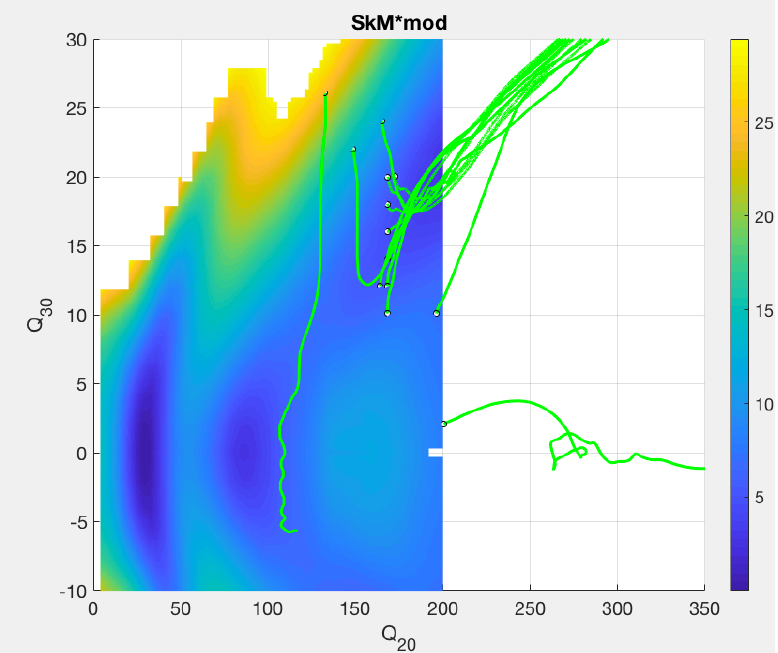}\hspace{2cm}\includegraphics[width=.4\linewidth]{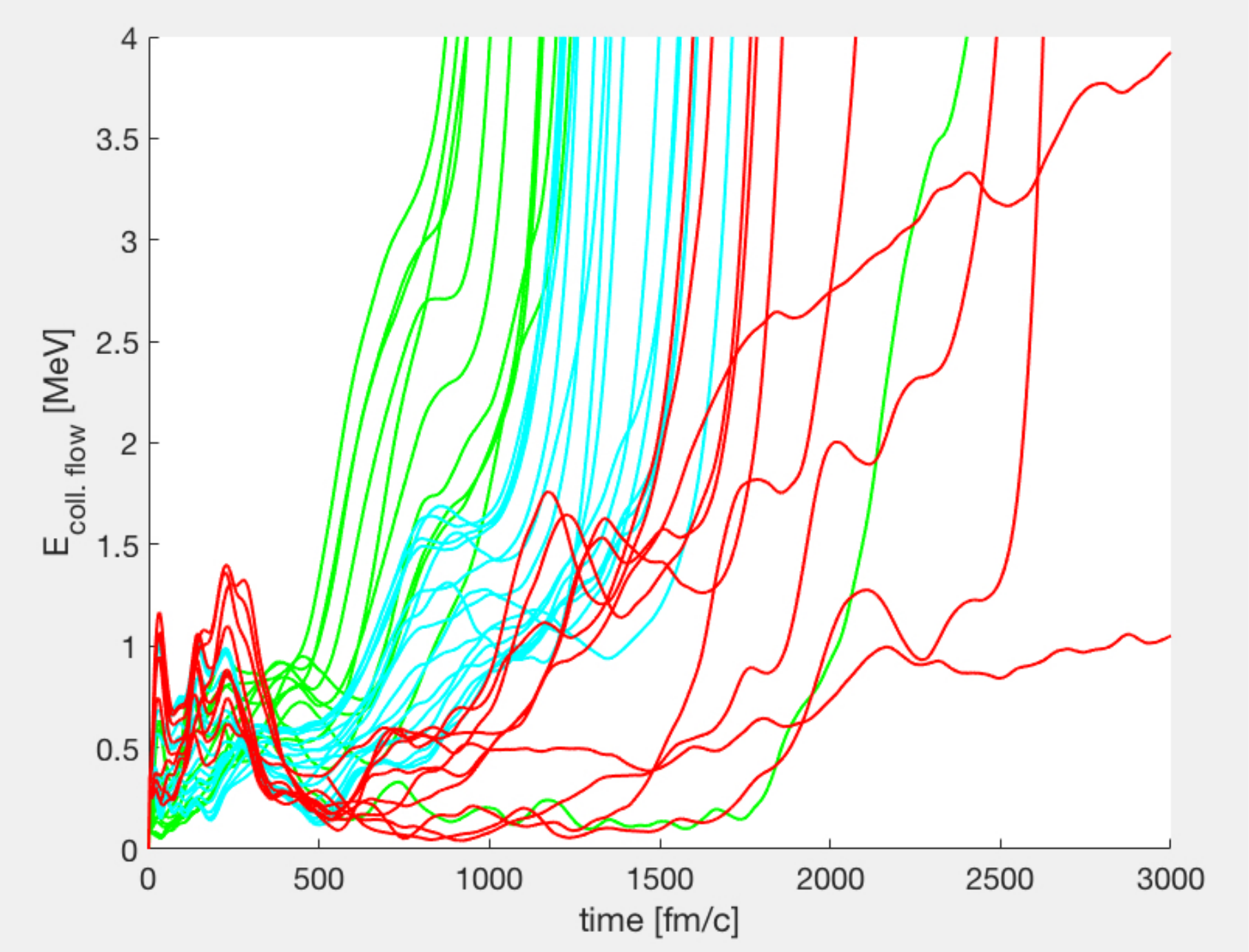}
\caption{Left: sample of trajectories from starting different points to fission configurations. Right: the collective flow energy remains very low during fission trajectories, up to the scission point.}
\label{fig:trajects}
\end{figure}

The initial state in TDSLDA simulation is located beyond the fission barrier, but close to the saddle point. If the initial state were placed in the ground-state potential well, even at an energy above the fission barrier, it would take a very long time to evolve toward a scission configuration, making the simulation impossible because numerical errors accumulate after a large number of time steps. Thus, a first valid question is how sensitive to the choice of the initial state our results are. In addition, the choice of the NEDF used as input in the calculations can influence the results. In Ref. \cite{Bulgac:2015a}, the choice was to use the generic Skyrme SLy4 parameterization for the NEDF, which is not considered one of  the appropriate parameterizations by fission practitioners, since it provides a very poor description of the potential energy surfaces,  in particular of the fission barriers. For our latest investigation presented in Ref. \cite{Bulgac:2018ukd}, we have used two more realistic NEDFs to calculate several trajectories, starting from different points on the potential energy surface. In the left hand panel of  Fig. \ref{fig:trajects}, we show several trajectories in the ($Q_{20}$, $Q_{30}$) space for the SKM* NEDF. The initial states have an average excitation energy of 8.3 MeV, with a standard deviation of 3 MeV, not including the symmetric trajectory (starting around $Q_{30}\approx0$).

Most of the trajectories shown in Fig. \ref{fig:trajects} produce fission fragments with very similar properties  $\langle N_l\rangle = 61.8(9)$ and $\langle Z_l\rangle=40.9(5)$, $\mathrm{TKE}\rangle = 174.5(2.5)$ MeV and $\langle \mathrm{TXE}\rangle = 31.5(3.8)$ MeV. These results are significantly different from the result that ends in the symmetric region, where TKE is much lower (149 MeV). One trajectory ends up in a local minimum and in the absence of fluctuations, it will take a long time for the system to fission. Although the initial conditions are all located on different points of the energy surface, the corresponding trajectories leading to asymmetric fission produce fragments with very similar properties, since the mean field by definition provides an average path toward fission. Fluctuations missing from the mean field are expected to play an important role to reach a good agreement of the simulations with experimental data on fission mass and TKE yields. Surprisingly, only a relatively narrow ensemble of initial conditions was considered in the TDHF+BCS calculations of Tanimura et. al. \cite{Tanimura:2017} resulted in distributions with widths comparable with experimental data. The simulations presented in this contribution and previously \cite{Bulgac:2018ukd} do not support their findings.
 
One particularly important feature of fission dynamics is illustrated in the left-hand side of Fig \ref{fig:trajects}, where we plot the collective flow energy as a function of time. The collective flow energy is defined as 
\begin{equation}
E_\mathrm{coll. flow} = \int d^3\vec{r}\frac{\vec{j}^2(\vec{r},t)}{2M_N\rho(\vec{r},t)},
\end{equation} 
where $\vec j(\vec r,t)=\frac{i\hbar}{2}\sum_k\left(v_k(\vec r,t)^*\vec\nabla v_k(\vec r,t)-v_k(\vec r,t)\vec\nabla v_k^*(\vec r,t)\right)$ is the current density, and $\rho(\vec r,t)=\sum_k|v_k(\vec r,t)|^2$ is the particle number density. For a point-like particle, this is simply the kinetic energy, and thus, from the classical point of view, if the particle is on an incline, one expects that the collective energy flow would increase quickly in time. Instead, we observe that the collective energy remains small (around 1 MeV) and almost constant, and it increases drastically only after scission, when the Coulomb repulsion takes over. This is in contrast with adiabatic approaches, where one expects a full conversion of the collective energy potential surface into a collective flow energy of about 15 to 20 MeV from saddle to scission. Hence, these results are consistent with the hypothesis of overdamped collective motion, as assumed in the work by Randrup et. al. \cite{PhysRevC.84.034613}. The motion is strongly dissipative due to the strong  one-body dissipation.

It is well accepted today that these prompt neutron and gamma emission (that is, neutron or gammas emitted before any $\beta$ decay toward stability) and angular and/or energetic correlations between them can offer information regarding the fission process \cite{Talou2018}. For example, the average neutrons emitted as a function of mass can give indirect information regarding the energy sharing between fragments, since the most efficient way to de-excite above the neutron separation energy is by neutron emission. Hence, the higher the excitation energy the more neutrons on average are emitted from the fragments. In addition, it was observed that when the incident energy of the neutron inducing fission increases, the number of neutrons emitted from the heavy fragments increases, while the average number of neutrons emitted from the light fragment remains the same (see Fig. 6 in Ref. \cite{PhysRevC.34.218}). This is an experimental indication that the extra excitation energy goes mostly into the heavy fragment. No other microscopic models available today can predict this behavior. Only TDSLDA, where fragments can be fully separated and thus the excitation energy extracted, has the potential to make reliable predictions, for example by setting initial conditions at finite temperature. 
 
State-of-the-art phenomenological models that simulate the prompt neutron and gamma emissions \cite{PhysRevC.85.024608,becker2013,SCHMIDT2016107} rely on the assumption that the prompt particle emission takes place only after the particles are fully accelerated. Moreover, since the neutron emission is very fast, the mass yields are always measured after the neutron emission. Hence, corrections have to be applied in order to obtain information regarding quantities before neutron emission, which are used as input in simulations. From this point of view, additional corrections both for theory and experiment could be required if the number of neutrons emitted at scission and/or during the acceleration of the fission fragments is significant as suggested by some phenomenological models \cite{CARJAN2015178}. Our investigation suggests that  the average number of neutron emitted from scission to full acceleration can reach more than 0.4, almost independent of the trajectory \cite{Bulgac:2018ukd}. However, more simulations in bigger boxes are necessary in order to eliminate any possible numerical artifacts.
 
\section{Outlook}

We have presented evidence that the TDSLDA can be a very effective tool in answering qualitative and quantitative questions regarding the dynamics of the fission process. Our simulations suggest that the one-body dissipation is strong, which leads to an overdamped dynamics. As a consequence, the trajectories follow predominantly the direction of the steepest descent and it is expected that the fluctuations left out in the mean field would play an important role in describing  the widths of the distributions observed experimentally. Recently, we have formulated the framework to include such fluctuations and dissipation within the time-dependent DFT \cite{Bulgac:2018fluc}. While for testing purposes the theoretical modes was initially implemented in a simpler hydrodynamic approach that does not include pairing or shell effects, it is straightforward to extend the same framework in TDSLDA. In fact, we have already presented in Ref. \cite{Bulgac:2018fluc} a couple of TDSLDA trajectories including fluctuations and dissipation. While TDSLDA is significantly more demanding computationally than any other theoretical models for fission, a reasonable number of trajectories can be run with today's computing capabilities in order to obtain a reasonable distributions. 

In this contribution, the discussion was limited to a few aspects of fission dynamics, including the evidence for emission of neutrons during acceleration. However, because it can follow the dynamics of the fissioning system until full separation, TDSLDA has the unique ability to provide information on all fission observables, eventually as a function of the excitation energy of the fissioning nucleus. In this framework, one can study the excitation energy sharing mechanism between the two fragments, and, with some modification, the average spin of the fission fragments before neutron emission. This work is already planned and will be investigated as soon as computational resources become available.

\textbf{Acknowledgment.} Work for this contribution was supported by the U.S. Department of Energy through different programs (Nuclear Physics, SciDAC, Office of Science, ASC, and SSAA).


\bibliography{master,extra}

\end{document}